\documentclass[aps,prb,preprint,groupedaddress,showpacs]{revtex4-1}
\usepackage{graphicx}
\usepackage{helvet}
\usepackage{amsmath,amssymb}
\usepackage{bm}

\makeatletter
\newcommand{\Mtmdt}{$M$(tmdt)$_{2}$}
\newcommand{\Zntmdt}{Zn(tmdt)$_{2}$}
\newcommand{\Pdtmdt}{Pd(tmdt)$_{2}$}
\newcommand{\Nitmdt}{Ni(tmdt)$_{2}$}
\newcommand{\Pttmdt}{Pt(tmdt)$_{2}$}
\newcommand{\Autmdt}{Au(tmdt)$_{2}$}
\newcommand{\Cutmdt}{Cu(tmdt)$_{2}$}
\newcommand{\proton}{$^{1}$H}
\newcommand{\BC}{$^{13}$C}
\newcommand{\dpsigma}{$dp{\sigma}$}
\newcommand{\muB}{${\mu}_{\rm B}$}
\newcommand{\Tone}{$T_{1}^{-1}$}
\newcommand{\ToneT}{$(T_{1}T)^{-1}$}
\newcommand{\pTone}{$^{1}T_{1}^{-1}$}
\newcommand{\cTone}{$^{13}T_{1}^{-1}$}

\newcommand{\piTone}{$^{\pi}T_{1}^{-1}$}
\makeatother

\begin{document}

\title{Single-component molecular material hosting antiferromagnetic and spin-gapped Mott subsystems}
\author{Rina\ Takagi$^{1{\dagger}}$,\ Takamasa\ Hamai$^{1}$,\ Hiro\ Gangi$^{1}$,\ Kazuya\ Miyagawa$^{1}$,\ Biao\ Zhou$^{2}$,\ Akiko\ Kobayashi$^{2}$, and\ Kazushi\ Kanoda$^{1}$}
\affiliation{$^{1}$Department\ of\ Applied Physics,\ University\ of\ Tokyo,\ Bunkyo\ City,\ Tokyo,\ 113-8656,\ Japan \\
$^{2}$Department\ of\ Chemistry,\ College\ of\ Humanities\ and\ Sciences,\ Nihon\ University,\ Setagaya\ City,\ Tokyo,\ 156-8550,\ Japan}
 
\date{\today}

\begin{abstract}
We investigated a system based solely on a single molecular species, Cu(tmdt)$_{2}$, accommodating $d$ and ${\pi}$ orbitals within the molecule.
$^{13}$C nuclear magnetic resonance measurements captured singlet-triplet excitations of ${\pi}$ spins indicating the existence of a ${\pi}$-electron-based spin-gapped Mott insulating subsystem, which has been hidden by the large magnetic susceptibility exhibited by the $d$ spins forming antiferromagnetic chains.
The present results demonstrate a unique hybrid Mott insulator composed of antiferromagnetic and spin-singlet Mott subsystems with distinctive dimensionalities.
\end{abstract}

\maketitle

Strongly correlated electrons with orbital degrees of freedom exhibit remarkable phenomena of keen interest, as exemplified by colossal magnetoresistance and heavy fermion behavior \cite{1,2,3}.
Thus, finding new multi-orbital systems is expected to promote the discoveries of novel phenomena and properties.
It is well known that molecular conductors show various correlation-related phenomena, such as Mott transition \cite{4}, charge order/glass \cite{5,6}, spin liquid \cite{7} and so on.
However, most of the phenomena originate from a single type of orbitals, either HOMO (highest occupied molecular orbital) or LUMO (lowest unoccupied molecular orbital).

A family of materials, {\Mtmdt} (tmdt = trimethylenetetrathiafulvalenedithiolate), are newly emerging molecular systems solely composed of a single molecular species, in which a transition-metal ion, $M$, is coordinated by organic ligands, tmdt, from both sides \cite{8,9}, as shown in Figs. 1(a) and (b).
The molecular orbitals lying near the Fermi level, ${\epsilon}_{\rm F}$, are the $p{\pi}$ orbitals extended over the tmdt ligand and the {\dpsigma} orbital located around $M$ \cite{10,11}.
The energy-level difference between the $p{\pi}$ and {\dpsigma} orbitals can be controlled systematically by replacing $M$ (Fig. 1(c)).
For $M =$ Ni$^{2+}$ and Pt$^{2+}$, the {\dpsigma} orbitals are of higher energy than the $p{\pi}$ orbitals residing around ${\epsilon}_{\rm F}$ and the two electrons accommodated in the two $p{\pi}$ orbitals give $p{\pi}$ band semimetals with appreciable electron correlation \cite{12,13,14,15,16}.
Contrastingly, the {\dpsigma} orbital lies close to the two $p{\pi}$ orbitals in the $M =$ Cu$^{2+}$ system, which accommodates three electrons in the three orbitals \cite{11}.
Such \textit{intramolecular} multi-orbital nature with variable degeneracy differentiates {\Mtmdt} from the conventional charge-transfer salts and offers a novel platform for orbital-selective physics, as demonstrated in this paper.

{\Cutmdt} is insulating in resistivity and exhibits the magnetic susceptibility of Bonner-Fischer type with an exchange interaction $J_{d} = $ 169 K, as shown in Fig. 1(d).\cite{17}
A {\proton} nuclear magnetic resonance (NMR) study revealed the characteristics of the spin dynamics of the one-dimensional antiferromagnetic Heisenberg spins in the paramagnetic state and an antiferromagnetic ordering below $T_{\rm N} = 13$ K.\cite{18}
The band-structure studies suggest that the {\dpsigma} orbitals, which are arranged one-dimensionally (see Fig. 1(b)), form a quasi-one-dimensional (Q1D) band \cite{11,19}, indicating that the {\dpsigma} orbitals are responsible for a Q1D antiferromagnetic Mott insulator.
On the other hand, theoretical studies of a multi-orbital Hubbard model \cite{19,20} predict that {\Mtmdt} potentially hosts diverse magnetic and conducting states under the variations of the on-site Coulomb repulsions on the {\dpsigma} and $p{\pi}$ orbitals, $U_{dp{\sigma}}$ and $U_{p{\pi}}$, respectively.

In the present study, we use the NMR technique, which is capable of probing the constituent orbital selectively, to elucidate the orbital dependent properties in {\Cutmdt}, particularly focusing on the $p{\pi}$ orbital.
The experiments found that the $p{\pi}$ orbitals constitute a spin-gapped Mott insulating subsystem, the magnetism of which was overwhelmed by the {\dpsigma} spin component in the bulk magnetization measurement.
In addition, the electron correlation in the $p{\pi}$ subsystem is found to be exceptionally strong among organic materials.
This observation is consistent with the interpretation that {\Cutmdt} hosts a novel multi-orbital Mott insulating phase, in which very distinctive Mott insulating phases are coexistent, i.e., a Q1D antiferromagnetic subsystem and an anisotropic two-dimensional spin-gapped subsystem.
  
Selectively {\BC}-enriched {\Cutmdt} were synthesized in a similar way to the normal {\Cutmdt}, which was reported previously \cite{17}.
{\BC} NMR measurements were performed for the fine polycrystals of {\BC}-enriched {\Cutmdt} under a magnetic field of 8.00 Tesla.
The outers of the double-bonded carbons in the centre of tmdt were selectively enriched by {\BC} isotopes as shown in Fig. 1(a).
This site has a large hyperfine coupling with the $p{\pi}$-electron spins.
The spectra were obtained by the fast Fourier transformation of echo signals observed after the spin echo sequence, $({\pi}/2)_{x}-({\pi})_{x}$.

To characterize the orbitals carrying spins in {\Cutmdt}, we first compare the {\BC} NMR spectra of {\Cutmdt} and a $p{\pi}$-orbital system {\Nitmdt} at room temperature (Fig. 2(a)). 
The shift of NMR line (${\delta}$) arises from the chemical shift (${\sigma}$) and Knight shift ($K$) tensors (see Supplemental Material \cite{21}).
The ${\sigma}$ depends on the local chemical structure around the nuclear site, while $K$ reflects the spin susceptibility.
For the chemical shift, we referred to the spectrum of non-magnetic {\Zntmdt} (Fig. 2(a)),\cite{22,23} in which the Knight shift vanishes; thus, the isotropic chemical shift, ${\sigma}_{\rm iso} = 126$ ppm (shown as a gray dashed line in Fig. 2(a)), which is determined by the first moment of the spectrum, is used as the origin of the isotropic Knight shift $K_{\rm iso}$.
The $K_{\rm iso}$ of {\Cutmdt} is negative in contrast to the positive $K_{\rm iso}$ value in {\Nitmdt}.
This directly proves the opposite sign of the $^{13}$C hyperfine coupling constant for the two compounds, namely, spins in {\Cutmdt} and {\Nitmdt} are accommodated in different kind of orbitals.
Because {\Cutmdt} has narrower linewidth and larger spin susceptibility compared with {\Nitmdt},\cite{12,17} the former orbital should have a smaller anisotropic hyperfine coupling with the $^{13}$C nuclear spin than the latter.
According to the electron density calculation of {\Mtmdt},\cite{24} the {\dpsigma} orbital has much smaller amplitude at the {\BC} site (Fig. 1(a)) than that of the $p{\pi}$ orbital.
Therefore, the above experimental result indicates that spins in {\Cutmdt} mostly reside in the {\dpsigma} orbital (with negative isotropic hyperfine coupling) unlike {\Nitmdt} with spins accommodated in the $p{\pi}$ orbitals (with positive isotropic hyperfine coupling).

Figure 2(b) shows the temperature dependence of the {\BC} NMR spectra of {\Cutmdt}.
A tail extending above 200 ppm with a small fraction, which was sample-dependent, is considered to originate from unknown impurity phases; so only the colored regions in the spectra are used to evaluate the Knight shift.
As the temperature is decreased, the spectrum varies in its position and shape and is much broadened below 13 K, indicating an appearance of internal fields due to an antiferromagnetic transition (Fig. 2(b)).
The isotropic component of the Knight shift, $K_{\rm iso} (= {\delta}_{\rm iso} - {\sigma}_{\rm iso})$, in the paramagnetic state is plotted in Fig. 2(c).
$K_{\rm iso}$ shows a broad peak at approximately 100 K, which coincides with the maximum of the magnetic susceptibility.
If the magnetic susceptibility is normalized to $K_{\rm iso}$ at approximately the peak temperature, the scaling is satisfactory below the peak temperature, but fails at higher temperatures.
Because the magnetic susceptibility, at least its main contribution, is explained by the Q1D {\dpsigma} spins, the deviation implies a cancellation of $K_{\rm iso}$ due to an additional local field with positive $^{13}$C hyperfine coupling that is especially sensed at the {\BC} sites and develops with temperature.

{\BC} nuclear spin-lattice relaxation rate {\cTone}, which probes the dynamical spin susceptibility, is plotted in Fig. 3.
The relaxation rate was determined by fitting the relaxation curve by the stretched exponential function.
(For the procedure of determining the relaxation rate, see Supplemental Material \cite{21}.)
The {\cTone} shows a sharp peak at $T_{\rm N} = 13$ K and exhibits a broad minimum in the range of $30-40$ K, followed by a monotonous increase that persists up to room temperature.
We compare this behaviour with the previously reported {\proton} NMR {\Tone} ({\pTone}) in Fig. 3.\cite{18}
The temperature dependence of {\pTone} is consistent with the scaling theory for the one-dimensional $S = 1/2$ antiferromagnetic Heisenberg model (AFHM) \cite{Sachdev}, which has three characteristic regimes in the paramagnetic state outside of the critical region near $T_{\rm N}$: (i) a low-temperature regime of $20-50$ K dominated by staggered spin fluctuations, which result in a temperature-insensitive {\Tone}(a correction with logarithmic $T$-dependence might appear at $T << J_{d}$, if the magnetic ordering is pushed down to far lower temperatures as in Sr$_{2}$CuO$_{3}$ \cite{Barzykin});
(ii) an intermediate-temperature regime of $50-200$ K dominated by uniform spin fluctuations (see Supplemental Material \cite{21} for a scaling in this regime \cite{Bourbonnais,Wzietek}), which result in a linear temperature dependence up to the order of the temperature of exchange interaction $J_{d}/k_{\rm B}$ (169 K in the present system);
(iii) a high-temperature regime above 200 K, where spin fluctuations with every wave numbers equally appear so that {\Tone} levels off.
Note that the formation of a sharp peak in both of $^{13}T_{1}^{-1}$ and $^{1}T_{1}^{-1}$ at 13 K indicates the three-dimensional order due to the finite inter-chain exchange interactions.
In Fig. 3, {\cTone} and {\pTone} are plotted to coincide with each other at the temperature, $J_{d}/k_{\rm B} =169$ K.
In regime (ii), the temperature variations of $^{1}T_{1}^{-1}$ and $^{13}T_{1}^{-1}$ are well scaled to each other.
In regime (i), {\pTone} is somewhat smaller in magnitude than {\cTone}, which can come from the difference in the form factor between the {\BC} and {\proton} sites; the former site is nearly exclusively hyperfine-coupled to the {\dpsigma} spins in the same molecule, whereas the latter site has appreciable hyperfine couplings to {\dpsigma} spins in adjacent molecules as well.
Thus, the antiferromagnetic fluctuations are filtered to some extent at the {\proton} sites \cite{18}.
Most remarkably, in the regime (iii) of $T > J_{d}/k_{\rm B}$, {\cTone} shows additional relaxation contribution that rapidly grows with temperature in contrast to the behaviors of 1D AFHM followed by {\pTone}.

The additional contribution to {\cTone}, namely a deviation of {\cTone} from the scaled {\pTone} value, which is approximated as 1.5 sec$^{-1}$ for temperatures of 200 to 300 K (Fig. 3), is also plotted in the figure.
It is evident that this additional contribution to {\Tone} is particularly apparent at the {\BC} sites, indicating that the $p{\pi}$ spins are likely responsible for the contribution because of much stronger coupling to the {\BC} sites than to the {\proton} sites.
This is consistent with the quasi-degenerate feature of the {\dpsigma} and $p{\pi}$ orbitals in {\Cutmdt}.\cite{11,19}
Because the additional relaxation rate (denoted by {\piTone} hereafter) appears only at high temperatures, the magnetic ground state of $p{\pi}$ orbitals should be non-magnetic; then, thermally activated paramagnetic spins are observed at high temperatures.
{\Cutmdt} carries two $p{\pi}$ electrons for two tmdt ligands in addition to one {\dpsigma} electron hosting the Q1D magnetism.
Considering that the tmdt ligand form a dimeric arrangement with that in an adjacent molecule analogous to the ${\beta}$-type configuration \cite{25} familiar for charge-transfer salts, two cases are conceivable for the $p{\pi}$ electronic states.
One case is a band insulator, for which the band comprising the bonding orbitals in the tmdt dimers is fully occupied and has an energy gap to an unoccupied antibonding band.
The other case is a dimerized Mott insulator, where one $p{\pi}$ electron is localized on a tmdt because of Coulomb interactions and the spins form singlets due to the dimerization.

In the case of the band insulator, the paramagnetism is due to quasiparticles being thermally activated to the upper band; thus, {\ToneT} exhibits a temperature variation of the Arrhenius type (Supplemental Material \cite{21}).
When we fit the experimental data of {\piTone} by the form,
\begin{equation}
(T_{1}T)^{-1} {\propto} {\exp}(-{\Delta}/T),
\end{equation}
a band gap of $2{\Delta} = 2600$ K (0.22 eV) is deduced from the activation plot of $(^{\pi}T_{1}T)^{-1}$ (the inset of Fig. 4(a)).
However, such a large gap does not comply with the band-structure calculation, which predicts a vanishingly small band gap \cite{11}.
Furthermore, the magnitude of {\piTone} is too large to explain quasiparticle excitations over the gap of 2600 K.
For example, at 300 K (the temperature of one-tenth of the gap energy), the density of thermally activated quasiparticles should be several orders of magnitude smaller than the quasiparticle density in {\Nitmdt}, which has the $p{\pi}$-band Fermi surfaces \cite{14}.
Nevertheless, the {\piTone} value for {\Cutmdt} at 300 K is ${\sim}$ 1.4 sec$^{-1}$, which is comparable to that for {\Nitmdt}, ${\sim}$ 3 sec$^{-1}$.
Therefore, the case of the $p{\pi}$ band insulator is ruled out.

In the case of the dimer Mott insulator, the additional relaxation originates from the singlet-triplet excitations over a spin gap of ${\Delta}_{s} = 1600$ K, which is deduced from the Arrhenius plot of {\piTone} (the main panel of Fig. 4(a)).
The hyperfine field at the {\BC} site from $S = 1/2$ localized $p{\pi}$ spins can be rather large compared to that from conducting $p{\pi}$ spins.
According to Ref. [19], the network of tmdt ligands is two-dimensional and the largest transfer integral between the tmdt ligands is $t_{\rm B}= 250$ meV (B bonds in Fig. 4(c)), which is approximately twice as large as the second largest one, $t_{\rm Q} = 134$ meV (Q bond in Fig. 4(c)), and seven times larger than the intramolecular one ($t_{\rm intra} = -35$ meV).
Thus, it is most likely that a spin singlet is formed in a dimer connected with the largest transfer integral, $t_{\rm B}$.

The NMR shift shows much stronger temperature dependence than the magnetic susceptibility above 150 K (Fig. 2(c)), as we discussed before.
The deviation of $K_{\rm iso}$ from the scaled magnetic susceptibility, ${\Delta}K_{\rm iso}$, is reasonably attributed to the singlet-triplet excitations of the $p{\pi}$ spins discussed above.
As shown in Fig. 4(b), the shift deviation multiplied by temperature, $T{\Delta}K_{\rm iso}$, approximately follows the activation type of temperature dependence, ${\exp}(-{\Delta}_{s}/T)$ with ${\Delta}_{s} = 1500$ K, which is in general agreement with the gap value determined from {\piTone}, of 1600 K (Fig. 4(a)).
The above-obtained energy gap for the singlet-triplet excitations, ${\Delta}_{s} = 1550$ K (the average of the gap values determined from the relaxation rate and shift) enables us to calculate the $p{\pi}$ spin susceptibility, $^{\pi}{\chi}$, by applying the singlet-triplet excitation model, which gives
\begin{equation}
\chi = \frac{2N_{\rm A}g^{2}{\mu}_{\rm B}^{2}}{3k_{\rm B}T} \frac{3e^{(-{\Delta}_{s}/T)}}{1+3e^{(-{\Delta}_{s}/T)}},
\end{equation}
where $N_{\rm A}$ is the Avogadro constant, $g$ is the g-factor, ${\mu}_{\rm B}$ is the Bohr magneton, and $k_{\rm B}$ is the Boltzmann constant.
Then, the magnetic susceptibility is estimated, e.g. at 280 K, to be $4.2{\times}10^{-5}$ emu/mol, which explains the deviation of magnetic susceptibility from that of the Q1D {\dpsigma} spins at the same temperature, $3.1{\times}10^{-5}$ emu/mol (Fig. 1(d)) \cite{17}.
Using the hyperfine coupling constant of the $p{\pi}$ spins, we can also estimate $^{\pi}{\chi}$.
Employing $^{\pi}a_{\rm iso} = 4500$ Oe/({\muB} tmdt), an average of the $^{\pi}a_{\rm iso}$ values for the isostructural compounds {\Nitmdt} and {\Pttmdt},\cite{16} the spin susceptibility given by $^{\pi}{\chi}$ = $^{\pi}a_{\rm iso} {\Delta}K_{\rm iso}$ is determined to be $4.3{\times}10^{-5}$ emu/mol, which is in good agreement with the former two values.
The spectral narrowing at higher temperatures in {\Cutmdt} (Fig. 2(b)) is also explained by the activated $p{\pi}$ spins because the anisotropy in the Knight shift compensates that in chemical shift (Supplemental Material \cite{21}).
Thus, the NMR at the {\BC} sites captures the ${\pi}$-electron-based spin-gapped Mott insulator, which has been hidden behind the large {\dpsigma} spins, and demonstrates that {\Cutmdt} is a multi-orbital Mott insulator composed of two Mott subsystems with distinctive characters in magnetism and dimensionality.

The present results provide insight into the nature of the $p{\pi}$ Mott insulating phase.
A $p{\pi}$ spin model based on Fig. 4(c) is depicted in Fig. 4(d).
The bond thickness is drawn so as to be proportional to the exchange interaction, $J_{i}$ ($i =$ A, B and Q).
Assuming the Heisenberg type of spin coupling, the exchange interaction is given by $J_{i} = 4t_{i}^{2}/U_{p{\pi}}$, where $U_{p{\pi}}$ describes the on-site (tmdt) Coulomb repulsive energy.
Most simply, the energy gap for the singlet-triplet excitations, ${\Delta}_{s}$, corresponds to the largest exchange interaction, $J_{\rm B} = 4t_{\rm B}^{2}/U_{p{\pi}}$.
Then, the values of ${\Delta}_{s} = 1550$ K (the average of the gap values determined from the relaxation rate and shift) and $t_{\rm B} = 250$ meV give an estimate, $U_{p{\pi}} = 1.9$ eV, which is considerably higher than the $U$ values in conventional molecular conductors (typically 1 eV or less).
Taking the second largest exchange coupling, $J_{\rm Q} = 4t_{\rm Q}^{2}/U_{p{\pi}}$, the spin system is modeled to one-dimensional spin chains with $J_{\rm B}$-$J_{\rm Q}$ alternate exchange couplings and the spin gap is given by the form of $J_{\rm B}-J_{\rm Q}/2$ according to a theoretical treatment valid for $J_{\rm B} >> J_{\rm Q}$, \cite{26} resulting in $U_{p{\pi}} = 1.6$ eV.
The large $U_{p\pi}$ value is an indication of strong electron correlation in the $p{\pi}$ Mott phase and is explained as follows.
The conventional Mott insulators in charge-transfer salts have dimeric molecular structures \cite{27}, in which a dimer plays a role of one lattice site and accommodates a hole on average.
In this case, the effective on-site (dimer) $U$ is reduced from the original $U$ by the spatial extension of the dimer molecular orbital.
The $1:1$ salts would have no such reduction in the $U$ value but maintain highly correlated nature.
The tmdt subsystem in {\Cutmdt} is in such a situation because one tmdt accommodates one carrier; i.e. the highly correlated $1:1$ salt is embedded in {\Cutmdt}.
This feature is common to other types of {\Mtmdt}, which is thus expected to generally possess the highly correlated $p{\pi}$ electrons.
Noticeably, {\Autmdt} and {\Pdtmdt} show antiferromagnetic orders at approximately 100 K, which is an exceptionally high transition temperature for molecular conductors \cite{28,29}.
 
The novel hybrid Mott insulator with intramolecular orbital-selectivity substantiated in the present study provides a perspective that {\Cutmdt} can be a unique platform for the orbital-selective Mott transition (OSMT).
Because the OSMT is a key concept that underlies various attractive phenomena in multiband materials such as iron-based superconductors and heavy fermion systems, theoretical and experimental investigations in quest of novel OSMT candidates have been intensive \cite{30,31,32,33,34,35,36,37,38,39,40,41,42,43,44,45}.
The quasi-one-dimensional antiferromagnetic insulator and the anisotropic two-dimensional spin-gapped insulator residing in {\Cutmdt} are likely to show distinctive characters in the Mott transitions and the resultant metallic phases.
Synergetic phases that may emerge in sequence during the orbital-selective Mott transitions by pressure are of particular interest.

The authors thank S. Ishibashi and H. Seo for fruitful discussions.
This work was supported by the JSPS Grant-in-Aids for Scientific Research (S) (Grant No. 25220709), for Challenging Exploratory Research (Grant No. 24654101), and by the JSPS Fellows (Grant No. 13J03087).

\newpage

\begin{figure}
\begin{center}
\includegraphics[width=12.6cm,keepaspectratio]{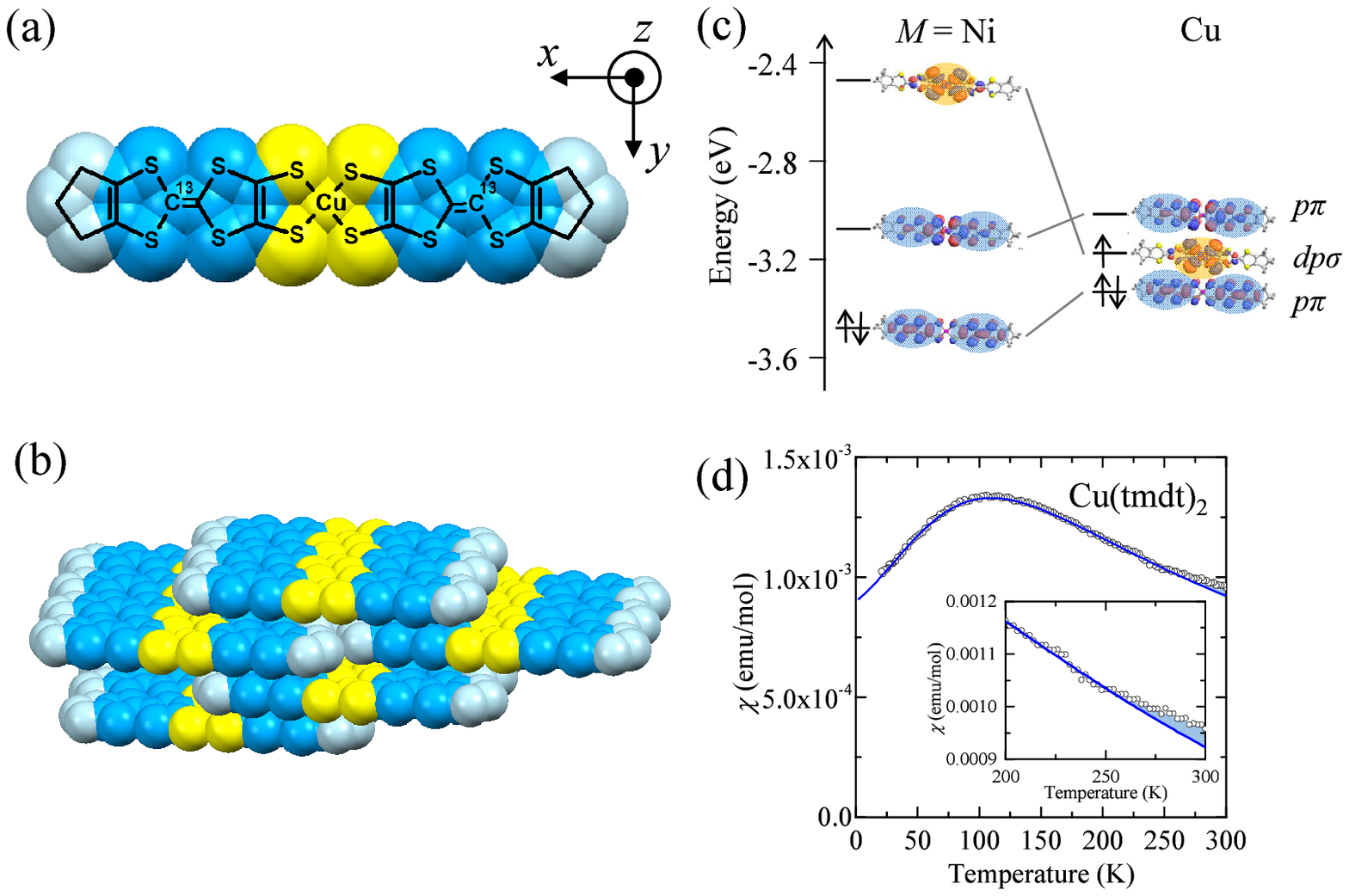}
\end{center}
\caption{(Color online) Structural, orbital and magnetic properties of {\Cutmdt}.
(a) Molecular structure of {\Cutmdt}.
Selectively-enriched {\BC} isotopes are labeled in the outers of the double-bonded carbons in the centre of the tmdt ligands.
(b) Crystal structure of {\Cutmdt}.
The {\dpsigma} orbital is populated on the central CuS$_{4}$ (yellow-colored part) and the $p{\pi}$ orbital is mainly populated in the blue-colored region in the tmdt ligand.
(c) Energy levels of the $p{\pi}$ and {\dpsigma} orbitals in {\Nitmdt} and {\Cutmdt}. \cite{10,11}
(d) Temperature dependence of the spin susceptibility, which is obtained by subtracting a low-$T$ Curie term from the raw data \cite{17}.
The blue line indicates the Bonner-Fischer-type fitting curve.
The deviation of measured data from the fitted curves for $T > 250$ K is indicated as the blue region in both the main panel and the inset.
}
\label{Fig1}
\end{figure}

\begin{figure}
\begin{center}
\includegraphics[width=12.6cm,keepaspectratio]{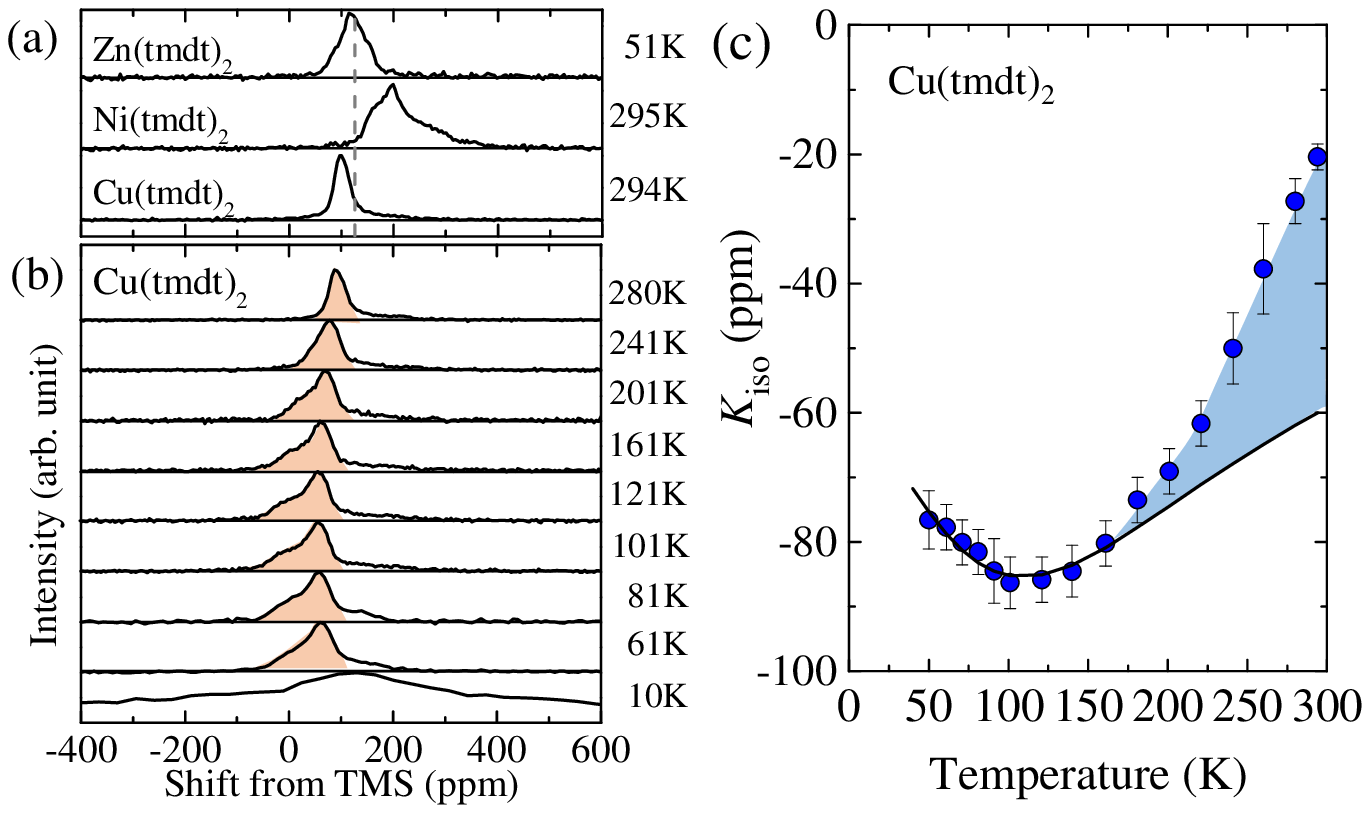}
\end{center}
\caption{(Color online) {\BC} nuclear magnetic resonance spectra and their analysis for {\Cutmdt}.
(a) {\BC} NMR spectra of {\Cutmdt} and the $p{\pi}$ electron conductor {\Nitmdt} at room temperature.
The {\BC} NMR spectrum of non-magnetic {\Zntmdt} at 51 K used for reference to the isotropic part of the chemical shift (a gray dashed line), ${\sigma}_{\rm iso}$ (= 126 ppm), is also shown.
(b) Temperature dependence of the {\BC} NMR spectra for {\Cutmdt}.
The colored regions in the spectra are used to determine the isotropic component of the Knight shift because tails without color are considered to originate from impurity phases.
(c) Temperature dependence of the isotropic part of the {\BC} NMR shift, $K_{\rm iso}$.
The value of $K_{\rm iso}$ is determined by subtracting ${\sigma}_{\rm iso}$ from that of the measured NMR shift, ${\delta}_{\rm iso}$.
The error bar is defined by the inhomogeneous width obtained in the spectral fitting (see Supplemental Material \cite{21}).
The solid curve is the magnetic susceptibility scaled to the shift values for $T < 140$ K.
The colored region indicates the deviation of $K_{\rm iso}$ from the scaled magnetic susceptibility, ${\Delta}K_{\rm iso}$, for temperatures above 150 K, which can be attributed to the singlet-triplet excitations of the $p{\pi}$ spins.
}
\label{Fig2}
\end{figure}

\begin{figure}
\begin{center}
\includegraphics[width=8.6cm,keepaspectratio]{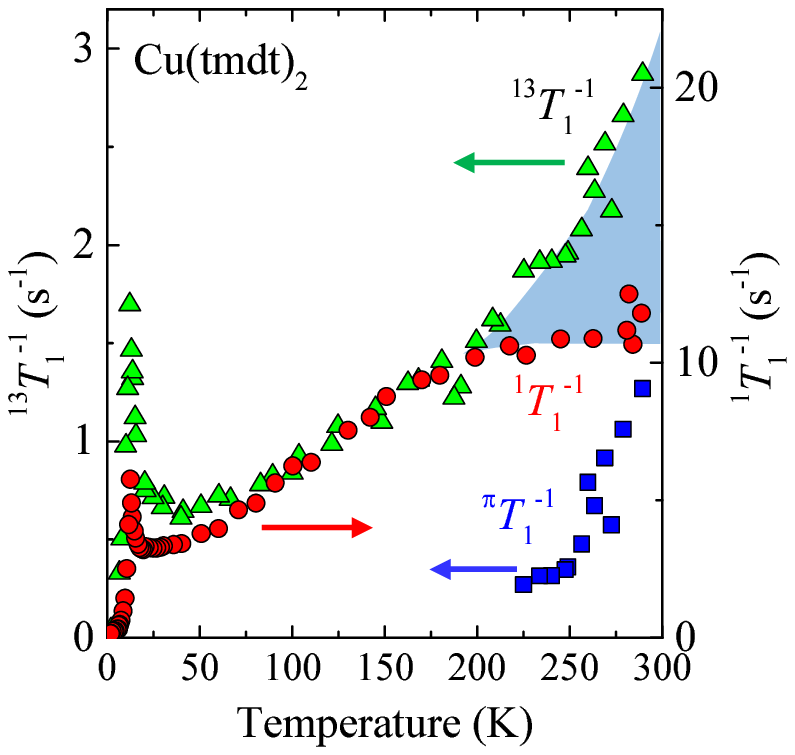}
\end{center}
\caption{(Color online) Nuclear spin-lattice relaxation rate, {\Tone}, for {\Cutmdt}.
The green triangles and the red circles indicate {\Tone} at the {\BC} and {\proton} sites (denoted as {\cTone} and {\pTone}, respectively, in the manuscript).
{\pTone} data are from Ref. [18].
{\cTone} and {\pTone} are plotted so that they coincide with each other at $T = J_{d}$ ($= 169$ K).
The deviation of {\cTone} from {\pTone} normalized to {\cTone} at $T = J_{d}$, (denoted as {\piTone} in the text) for temperatures above 200 K, represented by the colored region, is plotted using blue squares.
}
\label{Fig3}
\end{figure}

\begin{figure}
\begin{center}
\includegraphics[width=12.6cm,keepaspectratio]{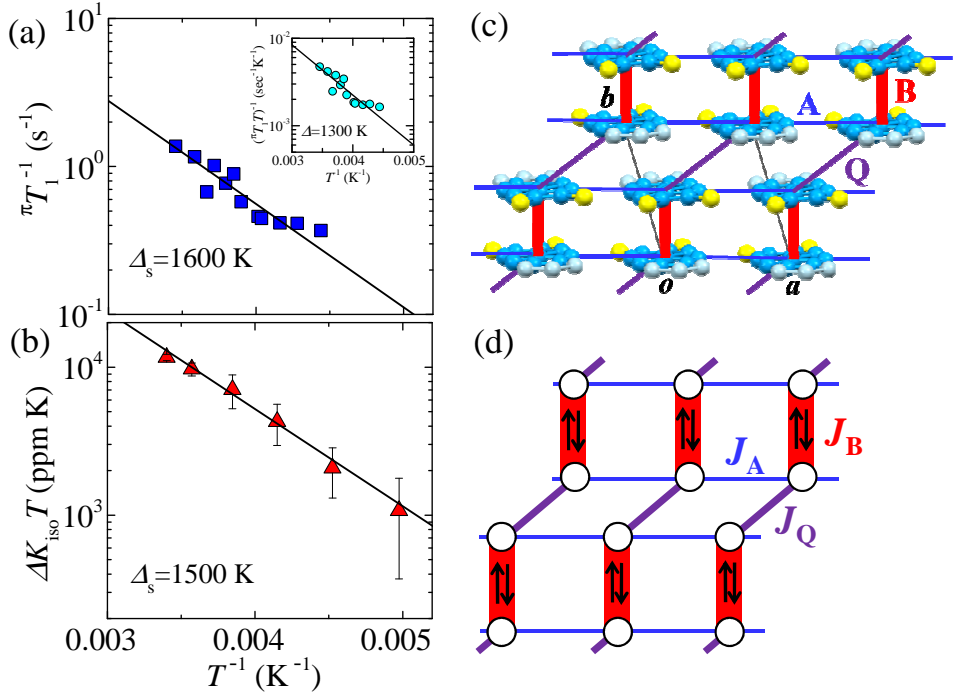}
\end{center}
\caption{(Color online) NMR evidence for a spin gap and its structural reasoning for {\Cutmdt}.
Activation plots for the $p{\pi}$-spin part of (a) {\BC} relaxation rate, {\piTone}, and (b) the isotropic component of {\BC} NMR shift multiplied by temperature, ${\Delta}K_{\rm iso}T$.
The inset of (a) is the activation plot of $(^{\pi}T_{1}T)^{-1}$.
(c) The two-dimensional network of tmdt ligands in the $ab$ plane.
Bonds between the tmdt ligands with significant transfer integrals are indicated as A: $[100]$, B: $[111]$, and Q: $[001]$; the transfer integrals are $t_{\rm A}= -90$ meV, $t_{\rm B}= 250$ meV and $t_{\rm Q} = 134$ meV.\cite{19}
The thickness of the bonding lines is proportional to the values of the transfer integrals.
(d) Spin model for the $p{\pi}$ electrons localized on the tmdt ligands corresponding to (c).
The bonding lines are drawn such that their thickness is proportional to the exchange interaction $J_{i} = 4t_{i}^{2}/U_{p{\pi}}$ ($i =$ A, B and Q) with the on-site Coulomb repulsive energy $U_{p{\pi}}$.
}
\label{Fig4}
\end{figure}

\end{document}


\title{Supplemental Material: \\
Single-component molecular material hosting antiferromagnetic and spin-gapped Mott subsystems}
\author{Rina\ Takagi$^{1{\dagger}}$,\ Takamasa\ Hamai$^{1}$,\ Hiro\ Gangi$^{1}$,\ Kazuya\ Miyagawa$^{1}$,\ Biao\ Zhou$^{2}$,\ Akiko\ Kobayashi$^{2}$, and\ Kazushi\ Kanoda$^{1}$}
\affiliation{$^{1}$Department\ of\ Applied Physics,\ University\ of\ Tokyo,\ Bunkyo\ City,\ Tokyo,\ 113-8656,\ Japan \\
$^{2}$Department\ of\ Chemistry,\ College\ of\ Humanities\ and\ Sciences,\ Nihon\ University,\ Setagaya\ City,\ Tokyo,\ 156-8550,\ Japan\\
$^{\dagger}$Current address: RIKEN Center for Emergent Matter Science (CEMS), Wako, Saitama 351-0198, Japan
}

\maketitle

\section{Determination of $^{13}$C nuclear spin-lattice relaxation rate}
 The local field at the {\BC} sites is distributed from grain to grain, depending on the geometry of the field direction against the crystal axes, because the {\BC} nuclear spins have anisotropic hyperfine coupling with the $p{\pi}$ electrons. Consequently, the nuclear relaxation rate {\Tone} is also different from grain to grain; thus, the nuclear relaxation curves in the polycrystalline sample are non-single exponential functions of time. We deduced the nuclear spin-lattice relaxation rate {\Tone} by fitting the nuclear relaxation curves to the stretched exponential function, $M(t) = M({\infty})[1- {\exp}(-(t/T_{1})^{\beta})]$, where $M(t)$ is the nuclear magnetisation at a time, $t$, after its saturation caused by the so-called rf comb pulse. The exponent ${\beta}$ characterizes the degree of distribution in $T_{1}$. As shown in Fig. S1, the ${\beta}$ values are in a range of $0.7-0.9$ at temperatures above $T_{\rm N}$ = 13 K. A sudden decrease in ${\beta}$ below $T_{\rm N}$ indicates an emergence of inhomogeneous local fields due to the magnetic ordering.

\section{Scaling relation between nuclear spin-lattice relaxation and magnetic susceptibility} 
As mentioned in the main text, $^{1}$H nuclear spin-lattice relaxation rate, $^{1}T_{1}^{-1}$, has three characteristic regimes in the paramagnetic state, among which the intermediate-temperature regime of $50-200$ K is dominated by uniform spin fluctuations.
In this regime, the temperature dependence of $^{1}T_{1}^{-1}$ is similar to the relaxation rate at $^{13}$C nuclear site, $^{13}T_{1}^{-1}$.
In case that one-dimensional uniform magnetic excitations dominate the nuclear relaxation rate, a scaling relation between $T_{1}^{-1}$ and the square of static spin susceptibility multiplied by $T$, $T{\chi}^{2}$, is suggested to hold [27].
Figure S2 shows $^{1}T_{1}^{-1}$ and $^{13}T_{1}^{-1}$ as a function of $T{\chi}^{2}$; both vary linearly with $T{\chi}^{2}$ in the temperature range of $50-130$ K and $50-150$ K, respectively, consistent with the scaling relation.
Here, we use the Bonner-Fischer-type fitting values as ${\chi}$.
We also tested the two-dimensional scaling relation, $T_{1}^{-1} \propto T{\chi}^{1.5}$.
As exhibited in the inset of Fig. S2, the scaling of $T_{1}^{-1}$ as a function of $T{\chi}^{1.5}$ does not better the linearity in the intermediate-temperature regime, reflecting the one-dimensional character of the {\dpsigma} spin system.

The extrapolation of the fitting line of $T_{1}^{-1}$ vs $T{\chi}^{2}$ to $T{\chi}^{2}=0$ gives a nonzero value of $T_{1}^{-1}$.
This indicates a sizable contribution of the staggered spin fluctuations to $T_{1}^{-1}$, which exceeds that of the uniform-spin fluctuations at lower temperatures.
The deviation of the data from the scaling line can be a measure of the relative contributions of the uniform and staggered spin correlations.

\section{Spin diffusion effect on the frequency dependence of the NMR relaxation rate}
Spin diffusion in quasi-one-dimensional systems gives a ${\omega}^{-1/2}$ dependence of $T_{1}^{-1}$ in particular for the long wavelength ($q$ ${\sim}$ 0) contributions.
In the case of {\Cutmdt}, uniform spin fluctuations are dominant for the temperature range of $50-200$ K (region (ii)), while spin fluctuations with every wave numbers equally contribute to $T_{1}^{-1}$ above 200 K (region (iii)).
Then, the spin diffusion effect would be more remarkable in the region (ii) than in the region (iii).
The observation frequencies for $^{13}T_{1}^{-1}$ and $^{1}T_{1}^{-1}$ are 86 and 156 MHz, respectively.
In the spin diffusion scenario, when $^{13}T_{1}^{-1}$ and $^{1}T_{1}^{-1}$ are plotted so as to coincide with each other at the temperature of $J_{d}/k_{\rm B} = 169$ K, which is around the boundary of the regions (ii) and (iii), $^{13}T_{1}^{-1}$ would deviate upward from $^{1}T_{1}^{-1}$ on cooling in the region (ii) while the deviation would be less prominent in the region (iii).
It is obvious in Fig. 3 of the main text that this is not the case; the temperature variations of $^{13}T_{1}^{-1}$ and $^{1}T_{1}^{-1}$ nearly the same in the region (ii) but largely deviate in the region (iii), as opposed to the expectation of the diffusion process.
Thus, the difference in the temperature dependence of $^{13}T_{1}^{-1}$ and $^{1}T_{1}^{-1}$ is not attributable to the frequency dependence of the NMR relaxation rate.

\section{Temperature dependence of the relaxation rate for a band insulator}
 In case of a band insulator, paramagnetism is carried by the quasiparticles thermally activated to the upper empty band from the lower occupied band near ${\epsilon}_{\rm F}$. For simplicity, we assume the relaxation due to s band electrons that occurs through the Fermi contact interaction,
\begin{equation}
H_{\rm F} = \frac{8\pi}{3} {\gamma}_{\rm n} {\gamma}_{\rm e} {\hbar}^{2} {\delta}(r){I_{z}S_{z}+\frac{1}{2}(I_{+} S_{-}+I_{-} S_{+} )},
\end{equation}
where ${\hbar}$ is the reduced Planck constant, $I_{z}$ and $S_{z}$ are the $z$-axis components of nuclear spin and electron spin, respectively, with the magnetic field applied in the $z$ direction. As is well known, the second and third terms cause the nuclear relaxation through simultaneous flipping of nuclear and electron spins accompanied by a momentum changes on scattering from ${\ket{\bvec{k}}}=u_{\bvec{k}}(\bvec{r}) e^{i{\bvec{k}}{\cdot}{\bvec{r}}}$ to ${\ket{\bvec{k^{'}}}} = u_{\bvec{k^{'}}}(\bvec{r})e^{i{\bvec{k^{'}}}{\cdot}{\bvec{r}}}$. If a band insulator with a gap ${\Delta}$ and a rectangular density of states is assumed for simplicity, then the transition probability $W$ is calculated as follows,

\begin{align}
W &{\simeq} \sum_{\bvec{k},\bvec{k^{'}}} \frac{2{\pi}}{\hbar}{\Bigl(}\frac{8{\pi}}{3} {\gamma}_{\rm n} {\gamma}_{\rm e} {\hbar}^{2}{\Bigr)}^{2} |{\bra{\bvec{k}}}{\delta}(\bvec{r}){\ket{\bvec{k^{'}}}}|^{2}  \frac{1}{4} {\delta}(E_{k}-E_{k^{'}})\nonumber \\
&={\iint} \frac{2{\pi}}{\hbar} {\Bigl(}\frac{8{\pi}}{3} {\gamma}_{\rm n} {\gamma}_{\rm e} {\hbar}^{2}{\Bigr)}^{2} |{\bra{\bvec{k}}}{\delta}(\bvec{r}){\ket{\bvec{k^{'}}}}|^{2} \frac{1}{4} {\delta}(E_{k}-E_{k^{'}}) N(E_{k})N(E_{k^{'}})\nonumber \\
&\quad \times f(E_{k}){\bigl(}1-f(E_{k^{'}}){\bigr)} dE_{k} dE_{k^{'}}\nonumber \\
&{\simeq} \int_{{\epsilon}_{\rm F}+{\Delta}/2}^{\infty} \frac{2{\pi}}{\hbar} {\Bigl(}\frac{8{\pi}}{3} {\gamma}_{\rm n} {\gamma}_{\rm e} {\hbar}^{2}{\Bigr)}^{2} |u_{k}(0)|^{4}  \frac{1}{4} N(E_{k})^{2} {\exp}{\Bigl(}\frac{{\epsilon}_{\rm F}-E_{k}}{k_{\rm B}T}{\Bigr)} dE_{k} \nonumber \\
&= 2{\exp}{\bigl(}-{\frac{\Delta}{2k_{\rm B}T}}{\bigr)} \frac{|u_{k_{\rm BE}}(0)|^{4}}{|u_{k_{\rm F}}(0)|^{4}} \frac{N(E_{k_{\rm BE}})^{2}}{N(E_{k_{\rm F}})^{2}} W_{\rm Korringa} ,
\end{align}
where $|u_{k_{\rm BE}}(0)|^{2}$ is the square of the Bloch wave function at a nuclear site at the edge of the conduction band, $N(E_{k_{\rm BE}})$ is the density of states at the edge of the conduction band, and $W_{\rm Korringa}$ is the transition probability for the Pauli-paramagnetic metallic case. Next, we obtain the form of {\Tone} as
\begin{equation}
\frac{1}{T_{1}} = 2W = \frac{1}{T_{1 \rm Korringa}}{\times}{\Bigl(}2 {\exp}{\bigl(}-\frac{\Delta}{2k_{\rm B}T}{\bigr)} \frac{|u_{k_{\rm BE}}(0)|^{4}}{|u_{k_{\rm F}}(0)|^{4}} \frac{N(E_{k_{\rm BE}})^{2}}{N(E_{k_{\rm F}})^{2}}{\Bigr)}.
\end{equation}
The activation-type temperature dependence is invariant, even in the case of anisotropic hyperfine coupling, the details of which are incorporated into the prefactors.

\section{Shifts from spectral fitting}
 We conducted a fitting analysis of the measured {\BC} NMR spectra by using the chemical shift tensor of non-magnetic {\Zntmdt} [22,23]. The hyperfine field at the {\BC} site mainly consist of an isotropic core-polarisation field and a uniaxially symmetric dipole field from the on-site $p_{z}$ orbital. The spin shifts of the present systems are characterized by an isotropic part, $K_{\rm iso}$, and an anisotropic part, $K_{\rm aniso}$. The total-shift tensor is given by
\begin{equation}
\begin{pmatrix}
\delta_{xx} &0 &0\\
0 &{\delta}_{yy} &0\\
0 &0 &{\delta}_{zz}
\end{pmatrix}
= 126 + 
\begin{pmatrix}
47 &0 &0\\
0 &-4.4 &0\\
0 &0 &-42.8
\end{pmatrix}
+ K_{\rm iso} + K_{\rm aniso}
\begin{pmatrix}
-1 &0 &0\\
0 &-1 &0\\
0 &0 &2
\end{pmatrix}
,
\end{equation}
where the $x$ and $y$ axes are in the molecular plane and the $z$ axis is perpendicular to the plane, as shown in Fig. 1(a) of the main text. The former two terms express the chemical shifts, and the latter two terms express the Knight shifts. The total shift for a field described in spherical coordinates, $({\theta}, {\phi})$, is given by ${\delta}({\theta}, {\phi}) = K_{xx}{\sin}^{2}{\theta} {\cos}^{2}{\phi} + K_{yy}{\sin}^{2}{\theta} {\sin}^{2}{\phi} + K_{zz}{\cos}^{2}{\theta}$.
Its powder-distribution, $f({\delta})$, has parameters, $K_{\rm iso}$ and $K_{\rm aniso}$, which are obtained by fitting $f({\delta})$ to the experimental spectra. In reality, however, inevitable inhomogeneity causes additional broadening of the spectra. Thus, in the analysis, we incorporated the inhomogeneous broadening by convoluting $f({\delta})$ with a Lorentzian function of the following form
\begin{eqnarray}
F({\delta}) &=& {\int} f({\omega}) \frac{\Delta}{({\omega}-{\delta})^{2}+{\Delta}^{2}} d{\omega} \nonumber\\
&=& {\iint} \frac{\Delta}{(K_{xx}{\sin}^{2}{\theta} {\cos}^{2}{\phi} +K_{yy}{\sin}^{2}{\theta}{\sin}^{2}{\phi}+K_{zz}{\cos}^{2}{\theta}-{\delta})^{2}+{\Delta}^{2}} {\sin}{\theta} d{\theta}d{\phi},
\end{eqnarray}
where ${\Delta}$ characterizes the inhomogeneous width and is assumed to have a form of ${\Delta}^{2}={\Delta}_{0}^{2}+a{\omega}^{2}$, with the second term expressing the width dependent on the shift. Eq. (S5) uses four parameters ($K_{\rm iso}$, $K_{\rm aniso}$, ${\Delta}_{0}$ and $a$) to fit the measured spectra. The curves obtained by the fitting are shown as red lines in Fig. S3(a), and the temperature dependence of $K_{\rm iso}$ and $K_{\rm aniso}$ are plotted in Figs. S3(b) and (c), respectively. The obtained $K_{\rm iso}$ (Fig. S3(b)) is nearly perfectly reproduced by the result determined by the first moment (Fig. 2(c) of the main text). $K_{\rm aniso}$ also shows a similar temperature dependence, consistent with the notion that their deviations from the solid lines (the scaled Bonner-Fisher susceptibility of the {\dpsigma} spins), ${\Delta}K_{\rm iso}$ and ${\Delta}K_{\rm aniso}$, originate from the $p{\pi}$ spins. In fact, the ratio, ${\Delta}K_{\rm aniso}/{\Delta}K_{\rm iso}$, which is 32 ppm/40 ppm = 0.8 at room temperature, is close to the ratio of the anisotropic and isotropic parts of the $p{\pi}$ hyperfine coupling constant, $^{\pi}a_{\rm aniso}$/$^{\pi}a_{\rm iso}$, which is estimated at 0.78 for {\Nitmdt} and 0.81 for {\Pttmdt}, thus corroborating the $p{\pi}$-spin origin for the magnetic excitations emerging at high temperatures.

 The sign reversal in $K_{\rm aniso}$ appears curious; however, the sign reversal can arise from the following reasons. In the presence of the $p{\pi}$ spins, their dipole fields dominate the anisotropic Knight shift, whereas, in the absence of $p{\pi}$ spins at low temperatures, the {\dpsigma} spins become the main contributors to the anisotropic shift. Concomitantly, the symmetry axis of the hyperfine coupling tensor is expected to change from the $z$ axis (the $p_{z}$ axis) to the $x$ axis (directed to the Cu ion) as shown in Fig. 1(a) of the main text. Our calculations of the dipole hyperfine fields at the {\BC} site by using the electron density calculation results [24] yield $(-1700, -2100, 3800)$ and $(56, -33,-22)$ in unit of Oe/{\muB} from the $p{\pi}$ spin and the {\dpsigma} spin (distributed on CuS$_{4}$), respectively. The above analysis assumes the former uniaxial symmetry (of $p_{z}$ orbital) and thus, at low temperatures where the $p{\pi}$ spins decay to zero, the symmetry axis is altered compared to that in the latter, possibly giving the spurious negative sign in $K_{\rm aniso}$. Note, however, that the anisotropic hyperfine terms from the {\dpsigma} spin are two orders of magnitude smaller than those from the $p{\pi}$ spins. Another possible reason for the sign change is the difference in chemical shift between {\Cutmdt} and {\Zntmdt} used as a reference. The near coincidence of Fig. 2(c) of the main text and Fig. S2(b) indicates that ${\sigma}_{\rm iso}$ is nearly the same between the two materials, with the difference arising in the anisotropy. The activation plot of $T{\Delta}K_{\rm aniso}$ yields ${\Delta}_{s}$ = 1400 K, which differs little from the original estimation (${\Delta}_{s}$ = 1500 K).

\section{Hyperfine coupling constant of $^{13}$C with $dp{\sigma}$ spins}
 Below 100 K, where the $p{\pi}$ spin susceptibility nearly vanishes, the isotropic part of the Knight shift, $K_{\rm iso}$, in conjunction with the magnetic susceptibility, ${\chi}$, gives the hyperfine coupling constant of the {\BC} nuclei with the {\dpsigma} spin, $^{d}a_{\rm iso}$, in two ways. One is to evaluate $^{d}a_{\rm iso}$ by the ratio of the $K$ and ${\chi}$ values at the peak temperature and gives $^{d}a_{\rm iso}$ = -360 Oe/({\muB} {\dpsigma}). The alternative way is to determine $^{d}a_{\rm iso}$ by the slants of the $K$-${\chi}$ plot below the peak temperature and gives $^{d}a_{\rm iso}$ = -340 Oe/({\muB} {\dpsigma}). The two values of $^{d}a_{\rm iso}$ are in good agreement with each other. The opposite signs of $^{\pi}a_{\rm iso}$ and $^{d}a_{\rm iso}$ clearly show the distinctive core-polarisation mechanisms in hyperfine couplings. Note that the absolute values of $^{d}a_{\rm iso}$ is one order of magnitude smaller than that of $^{\pi}a_{\rm iso}$, reflecting the situation in which the {\BC} site resides further from the {\dpsigma} spin than from the $p{\pi}$ spin.

\newpage

\begin{figure}
\begin{center}
\includegraphics[width=8.6cm,keepaspectratio]{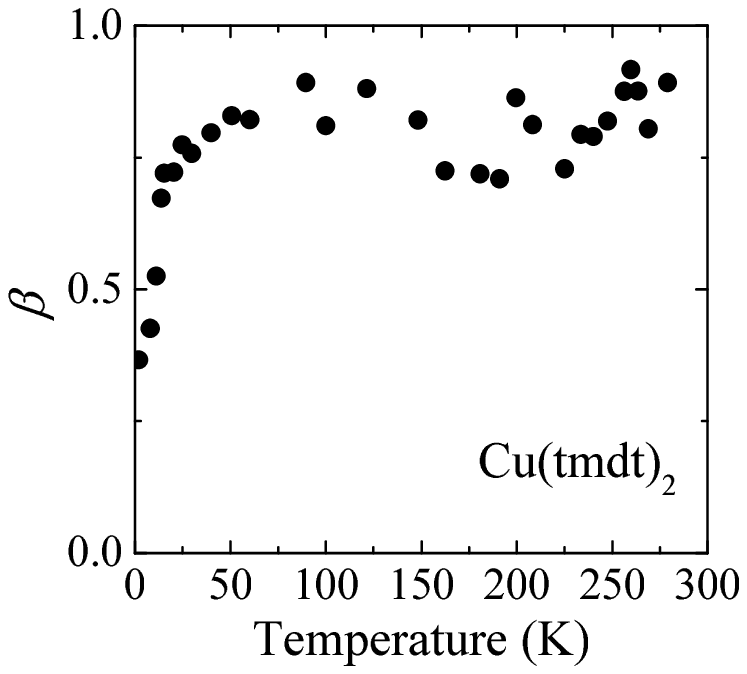}
\end{center}
\caption{(Color online) Temperature dependence of the exponent, ${\beta}$, in the stretched exponential fitting of the relaxation curve. The value of ${\beta}$ characterizes the degree of distribution in $T_{1}$. A decrease in ${\beta}$ below $T_{\rm N}$ indicates an emergence of inhomogeneous local fields due to magnetic ordering.}
\label{FigS1}
\end{figure}

\begin{figure}
\begin{center}
\includegraphics[width=12.6cm,keepaspectratio]{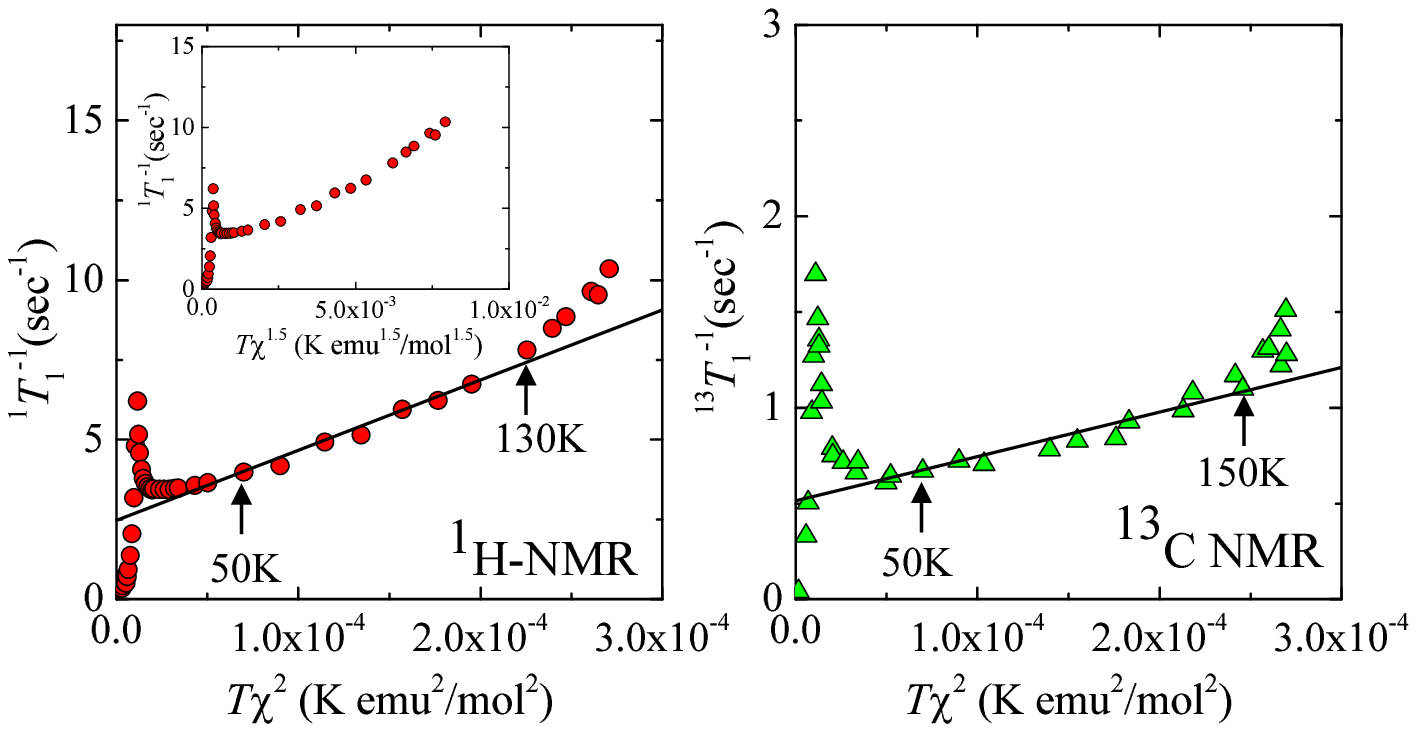}
\end{center}
\caption{(Color online) $^{1}$H and $^{13}$C nuclear spin-lattice relaxation rates, $^{1}T_{1}^{-1}$ and $^{13}T_{1}^{-1}$, with respect to $T{\chi}^{2}$ for Cu(tmdt)$_{2}$.
The solid lines are drawn based on the results of linear fitting of the data between $50-130$ K and $50-150$ K for $^{1}T_1^{-1}$ and $^{13}T_{1}^{-1}$, respectively.
The inset in the left panel shows $^{1}T_{1}^{-1}$ with respect to $T{\chi}^{1.5}$.}
\label{FigS2}
\end{figure}

\begin{figure}
\begin{center}
\includegraphics[width=12.6cm,keepaspectratio]{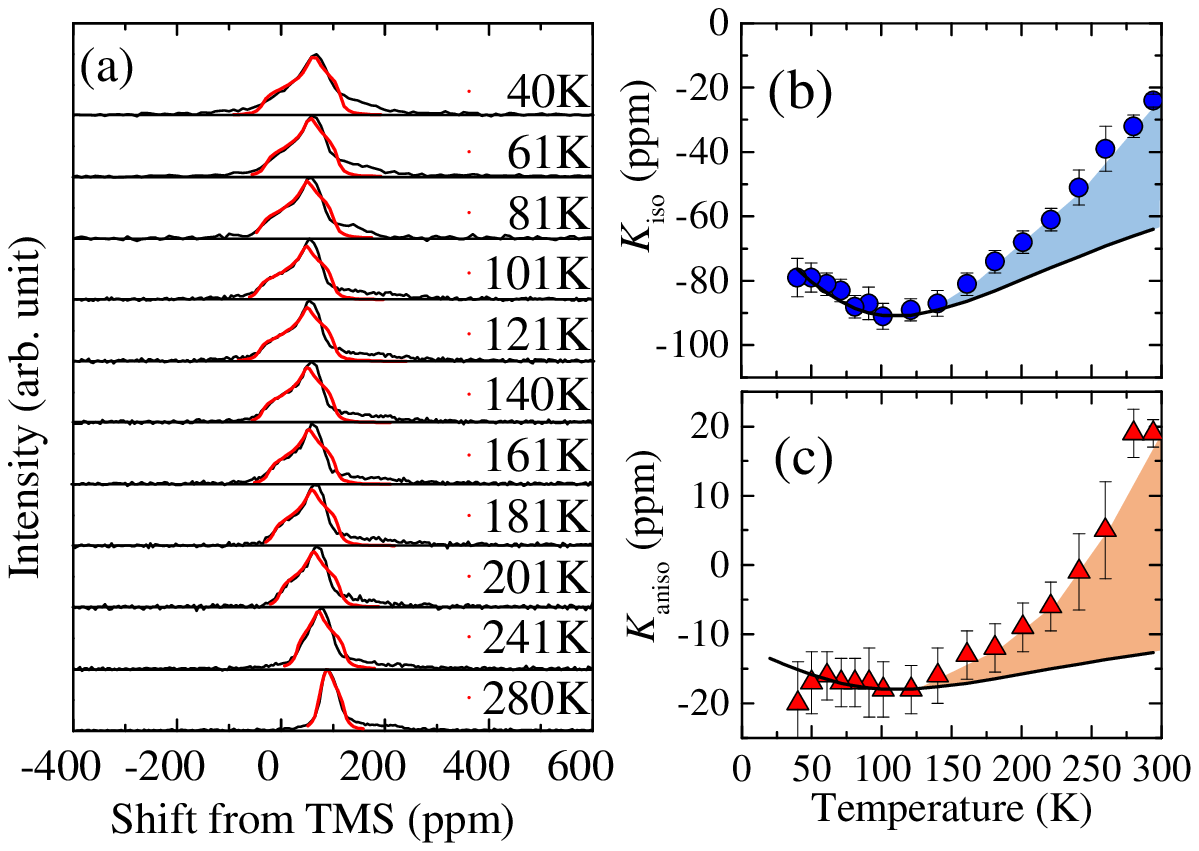}
\end{center}
\caption{(Color online) Determination of {\BC} NMR shift of {\Cutmdt}.
(a) Temperature dependence of {\BC} NMR spectra for {\Cutmdt}.
The black lines are the measured spectra and the red lines are fits described in the text.
(b), (c) Temperature dependences of the isotropic (b) and the anisotropic parts (c) of the {\BC} NMR shift deduced from the spectral fitting.
The error bars in (b) and (c) are defined by the inhomogeneous width obtained in the spectral fitting.
The black solid curves are the magnetic susceptibility scaled to the shift values at 100 K. The colored regions in (b) and (c) are the deviations of $K_{\rm iso}$ and $K_{\rm aniso}$, respectively, from the scaled magnetic susceptibility.}
\label{FigS3}
\end{figure}